\documentstyle[11pt]{article}
\def\mybaselinestretch{1.0}   
\newif\ifdodraftheader
\newif\ifdomarginnotes

\def\Titleofthispaper{%
        Vertex Models and Quantum Spin Systems: \\
        A nonlocal approach%
     \footnote{To appear in
               {\em Computer Simulations in Condensed Matter Physics VI},
                ed.\ D.P. Landau, K.K. Mon, and H.B. Sch\"uttler,
                Springer Verlag, Heidelberg, Berlin, 1993.}
       \\[0.5em]
}%
   \def\Preprintnumber{\rule{0ex}{1ex}FSU-SCRI-93C-65  \\[.5ex]     
                                      cond-mat/9305019 }           
   \def\Preprintdate  {May 1993}
\def\Authorsofthispaper{
     {\bf Hans Gerd Evertz$^{1}$}
     {\bf and Mihai Marcu$^{2}$}\\[2em]}
\def\Theiraddresses{
     $^1\,$Supercomputer Computations Research Institute,        \\[-.2ex]
           Florida State University, Tallahassee, FL 32306    \\[-.2ex]
           evertz@scri.fsu.edu                                   \\[3ex]
     $^2\,$School of Physics and Astronomy,\\[-.2ex]
           Raymond and Beverly Sackler Faculty of Exact Sciences,\\[-.2ex]
           Tel Aviv University, 69978 Tel Aviv, Israel           \\[-.2ex]
           marcu@taunivm.bitnet  \\[3ex]   
}%
\def\Abstracttext{%
Within a general cluster framework, we discuss
the loop-algorithm, a new type of cluster algorithm that
reduces critical slowing down in vertex models and in quantum
spin systems.
We cover the example of the 6-vertex model in detail.
For the F-model, we present numerical results
that demonstrate the effectiveness of the loop algorithm.
We discuss how to
modify the original algorithm for
some more complicated situations, especially for
quantum spin systems in one and two dimensions.
}
    \newcommand{\note}[1]{}
    \ifdomarginnotes
                      \renewcommand{\note}[1]
                             {\marginpar{\raggedright\tiny #1}
                              \message{Note: #1}}
              \fi
    \ifdodraftheader  
             \fi
 \let\section=\subsection
 \let\subsection=\subsubsection
 \def\subsubsection#1{\subsection{#1}
      \par\note{CAUTION: subsection=subsubsection !}\par}
\newcount\hour   \newcount\minute
\hour  = \time   \divide\hour by 60
\minute= \hour   \multiply\minute by -60   \advance\minute by \time
\ifnum\minute<10 \def\Time{\number\hour:0\number\minute}
           \else \def\Time{\number\hour:\number\minute}  \fi

\newcommand{\figurebox}[2]{\fbox{\vbox to #1{\hbox to #2{\hfil}\vfil}}}
\newcommand{\stru}[1]{\rule{0ex}{#1}}   
\newcommand{\str}{\rule{0ex}{2.7ex}}    
\newcommand{\tabhline}{\\[0.3ex] \hline \str}

\newcounter{itemnumber}

\let\mc=\multicolumn

\newcommand{\half}{{1\over2}}		    

\newcommand{\myeq}{\!=\!}

\newcommand{\mygt}{\!>\!}

\newcommand{\myleq}{\!\leq\!}

\def\3{\ss}

\def\kch{\half K_c}
\def\qmin{q_{\mbox{{\protect\scriptsize 3,min}}}}
\def\zcl{z^{\mbox{{\protect\scriptsize cl}}}}
\def\taucl{\tau^{\mbox{{\protect\scriptsize cl}}}}

\def\tauint{\tau_{\mbox{{\protect\scriptsize int}}}}

\def\zint{z_{\mbox{{\protect\scriptsize int}}}}

\def\zatKc{0.71(5)}
\def\zatKch{0.19(2)}
\newcommand{\U}{\tilde{u}}
\newcommand{\ovl}[1]{\overline{#1}}
\begin{document}
%
  \parbox[t]{20ex}{\ifdodraftheader \fbox{\bf Draft version}   \fi}
  \hfill           \ifdomarginnotes \note{\today} \note{\Time} \fi
  \begin{tabular}[t]{l}
                     \rule{0ex}{1ex}\Preprintnumber \\[.5ex]
                     \rule{0ex}{1ex}\Preprintdate
  \end{tabular}
%
%
    \renewcommand{\thefootnote}{{\protect\fnsymbol{footnote}}}
     \vskip 2em
         \begin{center}
           {\Large \bf \vbox{\vspace{4ex}} \Titleofthispaper \par}\vskip1.5em
         {\normalsize \lineskip .5em
             \begin{tabular}[t]{c} \vbox{\vspace{3em}} 
                                 \Authorsofthispaper
             \end{tabular}\par}
   {\normalsize \begin{tabular}[t]{c} \vbox{\vspace{0em}}
                                 \Theiraddresses
             \end{tabular}\par}  \vskip 1em
         \end{center} \par 
    \renewcommand{\thefootnote}{{\protect\arabic{footnote}}}
 \vfill
 \pagebreak[3]
\begin{abstract} {\protect\normalsize
                   \noindent \Abstracttext}
\end{abstract}
\vfill
\thispagestyle{empty}
\setcounter{page}{0}
\newpage
\renewcommand{\baselinestretch}{\mybaselinestretch} 
\protect\small \protect\normalsize                  
%
%
%
 \section{Introduction}
%
For Monte Carlo simulations of many interesting physical situations,
critical slowing down is a major problem.
Standard simulation algorithms employ {\em local} update procedures
like the Metropolis and the heat bath algorithm.
With local updates, ``information'' travels slowly, like a random
walk. If the
relevant length scale is the correlation length $\xi$,
the number of updates necessary to decorrelate large regions,
i.e.\ the autocorrelation time $\tau$, grows like
\begin{equation}\label{CSD}
\tau \propto \xi^z,
\end{equation}
where $z \approx 2$ for local updates, as suggested by the
random walk analogy.

The way out of this problem is to employ $nonlocal$ updates.
The challenge is to devise algorithms that are nonlocal and still
satisfy detailed balance.
Multigrid algorithms are one possible approach.
In this paper we shall focus on {\em cluster algorithms};
for a nonexhaustive selection of references see
\cite{SW,Wolff,ClusterReviews,KandelDomany,VMR,Z2gauge}.

The first cluster algorithm was invented by
Swendsen and Wang (SW) \cite{SW} for the case of the
Ising spin model. The basic idea is to perform moves that significantly
change the Peierls contours characterizing a configuration.
As the size of Peierls contours is, typically, anything up to the
order of the correlation length, critical slowing down may be eliminated
completely or at least partially by this approach.
The SW algorithm has been modified and generalized for other spin systems,
mostly with two spin interactions \cite{Wolff,ClusterReviews,VMR}.
Notice that for these systems clusters are connected regions of spins,
with the same dimensionality as the underlying lattice.
A few generalizations along different lines were also done
\cite{KandelDomany,Z2gauge}.

Recently \cite{BCSOScluster,LoopPRL,LoopSixVertex} we introduced
{\em cluster algorithms for vertex models and quantum spin systems},
which are  the first cluster
algorithms for models with constraints. While \cite{BCSOScluster} is an
adaptation of the valleys-to mountains-reflections (VMR) algorithm
\cite{VMR}, originally devised for solid-on-solid models,
the {\em loop algorithm} introduced in \cite{LoopPRL,LoopSixVertex}
does not resemble any existing scheme.

In this paper we discuss the loop algorithm in detail.
In vertex models \cite{ReviewLieb} the dynamical variables
are localized on bonds, and the interaction is between all bonds meeting
at a vertex. Furthermore there are constraints on the possible bond
variable values around a vertex.
Our scheme is devised such as to take into account the constraints
automatically, and to allow a simple way to construct the  clusters.
The clusters here are not connected regions of spins,
but instead {\em closed paths} (loops) of bonds.

In what follows we first comment on the SW algorithm for spin systems.
Then we discuss the general cluster formalism of Kandel and Domany
\cite{KandelDomany}, treating the SW algorithm as an example.
Next we define the 6-vertex model, and, as a special case, the F~model.
We then introduce the loop algorithm, and show how to formulate it
for the complete 6-vertex model. The optimization of the algorithm is
discussed in a separate section. For the special case of the F-model,
we describe how the algorithm particularizes, and then we present
our very successful numerical tests of the loop algorithm.
We also show how to apply the loop algorithm to simulations of one and two
dimensional quantum spin systems \cite{QMC},
like e.g. the $xxz$ quantum spin chain,
and the spin $1\over2$ Heisenberg antiferromagnet and ferromagnet
in two dimensions.
Finally we comment on further generalizations and applications.

 \section{Some comments on the Swendsen-Wang algorithm}

Here we present a somewhat unusual viewpoint on the SW algorithm, in
order to better understand what is new in the loop algorithm.
Let us look at a spin system, like the Ising model,
with variables $s_x$ living on sites $x$,
and a nearest neighbor Hamiltonian $H(s_x,s_y)$.
The partition function of the model to be simulated is
$Z=\sum_{s} \exp{(-\sum_{<xy>} H(s_x,s_y))}$, where $<xy>$ is a pair
of sites.

Consider update proposals (flips) $s_x \rightarrow s'_x$,
such that $H(s_x,s_y) = H(s'_x,s'_y)$.
Then it follows that updating at the same time
all spins in some ``cluster'' of spins, we
will only change the value of the Hamiltonian at the {\em boundary} of that
cluster, not inside it.

In order to satisfy detailed balance,
we have to choose clusters with an appropriate probability.
The SW algorithm amounts to making a {\em Metropolis} decision
for each bond $<xy>$, namely whether to accept the change in $H$ from
$H(s_x,s_y)$ to $H(s'_x,s_y)$.
If accepted, a bond is called ``deleted'', otherwise it is called
``frozen''. Clusters are then sets of sites connected by frozen bonds.
Note that if deleted, a bond {\em may} be at the boundary of a cluster,
but need not.

Finally, an update is performed by finding all the clusters in a given
configuration and then flipping each cluster with 50\% probability
\cite{SW}. In Wolff's single cluster variant \cite{ClusterReviews,Wolff},
which is dynamically more favourable,
we construct one cluster starting from a randomly chosen
initial site, and then flip it with 100\% probability.

A technical remark: The Swendsen-Wang algorithm can be
vectorized and parallelized \cite{ParallelSW}.
The difficult task is to identify
the clusters, which is the same task as e.g.\ image component labeling.
For the single cluster variant, vectorization is the most efficient
approach \cite{VectClust}.

 \section{The Kandel - Domany framework.}

Cluster algorithms are conveniently described in the general framework
of Kandel and Domany \cite{KandelDomany}.
Let us consider the partition function
\begin{equation}\label{ZKandel}
Z=\sum_{u} \exp{(-V(u))},
\end{equation}
where $u$ are the configurations to be summed over. We shall call
the function $V$ the ``interaction''.
Let us also define a {\em set} of new interactions $\ovl{V}_i$
(the index $i$ numbers the modified interactions).
Assume that during a Monte Carlo simulation we arrived at a given
configuration $u$. We choose a new configuration in a two step
procedure. The first step is to replace $V$ with one of the
$\ovl{V}_i$. For a given $i$, $\ovl{V}_i$ is chosen with
probability $p_i(u)$. The $p_i(u)$ satisfy:
\begin{equation} \label{KDconstraints}
\begin{array}{rll}
       p_i(u) &=& exp{(V(u) -  \ovl{V}_i(u) +c_i)}, \\[0.5ex]
\sum_i p_i(u) &=& 1 \; ,
\end{array}
\end{equation}
where $c_i$ are constants independent of $u$.
The second step is to update the configuration by employing a
procedure that satisfies detailed balance for the chosen $\ovl{V}_i$.
The combined procedure satisfies detailed balance for the
original interaction $V$ \cite{KandelDomany}.

In many cases, the interaction $V$ is a sum over local functions $H^c$,
where $c$ typically are cells of the lattice (like bonds in the
spin system case, sites for vertex models, plaquettes for gauge theories).
More generally, $H^c$ can contain part (or all) of the interactions in
some neighborhood of the cell $c$.
We can choose the modified interactions
$\ovl{H^c}_{i^c}$ for each $H^c$ independently ($i^c$ numbers the
possible new interactions for the cell $c$). The probabilities
for this choice obey eq.\ (\ref{KDconstraints}),
with $V$ replaced by
$H^c$, and $i$ by $i^c$. The total modified interaction
$\ovl{V}_i$ will then be the sum of all the $\ovl{H^c}_{i^c}$
($i$ is now a multiindex). When clear from the context, we shall
drop the index $c$.

Take the Ising model as an example.
The configuration $u$ is comprised of spins
$s_x = \pm 1$, the cells where the interaction is localized are
bonds $<xy>$, and $V(u) = -J \sum_{<xy>} s_x s_y$. We choose the bonds
as our cells $c$, so we can perform the first step of the
Kandel-Domany procedure separately for each bond.
The original cell interaction is
\begin{equation}
H^{<xy>}(s) = -J s_x s_y \;.
\end{equation}
Now, let us define two new bond interactions; the first ($i^{<xy>}=1$)
is called ``freeze'', the second ($i^{<xy>}=2$) is called ``delete'':
\begin{equation}\label{KDising}
%
\ovl{H^{<xy>}}_{\mbox{\small freeze}}(s) =
   \left\{ \begin{array}{ll}  0,  & s_x = s_y \\
            \infty, & else   \end{array}  \right. ; \quad\quad
\ovl{H^{<xy>}}_{\mbox{\small delete}}(s) = 0 \; .
\end{equation}
Then from (\ref{KDconstraints})
we obtain the Swendsen-Wang probabilities
if we choose the constants $c_i$ that minimize freezing:
\begin{equation}
\begin{array}{lll}
p^{<xy>}_{\mbox{\small delete}}(s) &=&
  \exp{(-J (s_x s_y + 1))} \, = \, \min{(1,\exp{(-2J s_x s_y)})} \;, \\[1.5ex]
p^{<xy>}_{\mbox{\small freeze}}(s) &=&
  1 - p^{<xy>}_{\mbox{\small delete}}(s) \;,
\end{array}
\end{equation}
which is just the Swendsen-Wang probability.

 \section{The 6-Vertex Model}

The 6-vertex model \cite{ReviewLieb} is defined on a square
lattice. On each bond there is an Ising-like variable that is usually
represented as an arrow. For example, arrow up or right means $+1$,
arrow down or left means $-1$. At each vertex we have the constraint
that there are two incoming and two outgoing arrows.
In fig.~1
we show the six possible configurations
at a vertex, numbered as in \cite{ReviewLieb}.
The statistical weight of a configuration is given by the product over
all vertices of the vertex weights $\rho(u)$. Thus a priori,
for each vertex there are 6 possible weights $\rho(u)$, $u=1,...,6$.
We take  the weights to be symmetric under
reversal of all arrows. Thus, in standard notation
\cite{ReviewLieb}, we have:
\begin{equation}\label{sixvertexweights}
\begin{array}{l}
  \rho(1)  =  \rho(2)  =  a \; , \\
  \rho(3)  =  \rho(4)  =  b \; , \\
  \rho(5)  =  \rho(6)  =  c \; .
\end{array}
\end{equation}
The 6-vertex model has two types of phase transitions: of
Kosterlitz-Thouless type and of KDP type \cite{ReviewLieb}.
A submodel exhibiting the former is the F~model, defined by
$c=1$, $a=b=\exp{(-K)}$, $K\geq0$.
For the latter transition an example is the KDP
model itself, defined by $a=1$, $b=c=\exp{(-K)}$, $K\geq0$.

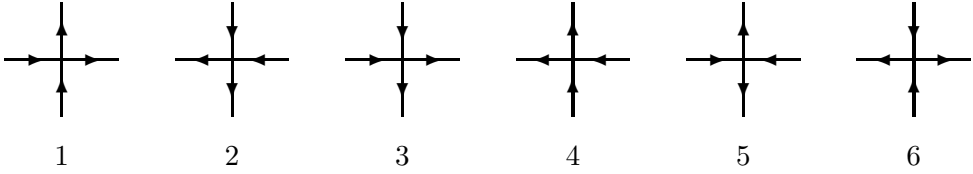
\begin{figure}[tb] \label{figsixvertices}
\begin{center}
\setlength{\unitlength}{.0005\textwidth}
\begin{picture}(1700,200)(0,50) 
\thicklines
%
\newsavebox{\RIGHT}
\newsavebox{\LEFT}
\newsavebox{\UP}
\newsavebox{\DOWN}
\sbox{\RIGHT}{\put(  0,0  ){\vector( 1, 0){70}}\put(70,0 ){\line(1,0){30}}}
\sbox{\UP}   {\put(  0,0  ){\vector( 0, 1){70}}\put(0 ,70){\line(0,1){30}}}
\sbox{\LEFT} {\put(100,0  ){\vector(-1, 0){70}}\put(0 ,0 ){\line(1,0){30}}}
\sbox{\DOWN} {\put(  0,100){\vector( 0,-1){70}}\put(0 ,0 ){\line(0,1){30}}}
\newcommand{\VERTEX}[5]{\begin{picture}(0,0)
                              \put(-100,0   ){\usebox{#1}}       
                              \put(   0,0   ){\usebox{#2}}       
                              \put(   0,-100){\usebox{#3}}       
                              \put(   0,0   ){\usebox{#4}}       
                              \put(   0,-170){\makebox(0,0){#5}} 
                        \end{picture}}
%
\put( 100,200){\VERTEX{\RIGHT}{\RIGHT}{\UP}  {\UP}  {1}}
\put( 400,200){\VERTEX{\LEFT} {\LEFT} {\DOWN}{\DOWN}{2}}
\put( 700,200){\VERTEX{\RIGHT}{\RIGHT}{\DOWN}{\DOWN}{3}}
\put(1000,200){\VERTEX{\LEFT} {\LEFT} {\UP}  {\UP}  {4}}
\put(1300,200){\VERTEX{\RIGHT}{\LEFT} {\DOWN}{\UP}  {5}}
\put(1600,200){\VERTEX{\LEFT} {\RIGHT}{\UP}  {\DOWN}{6}}
\end{picture}
\caption[fig1]{\parbox[t]{.80\textwidth}{
               The six vertex configurations, $u=1,...,6$
               (using the standard conventions of \cite{ReviewLieb}).}}
\end{center}
\vskip-1.5ex   
\end{figure}
%

\section{The Loop Algorithm}

If we regard the arrows on bonds as a vector field, the constraint at
the vertices is a {\em zero-divergence condition}. Therefore every
configuration change can be obtained as a sequence of {\em loop-flips}.
By ``loop'' we denote an oriented, closed,
non-branching (but possibly self-intersecting)
path of bonds, such that all arrows along the path point in the
direction of the path. A loop-flip reverses the direction  of
all arrows along the loop.

Our cluster algorithm performs precisely such operations, with
appropriate probabilities.
It constructs closed paths consisting of one or
several loops without common bonds. All loops in this path are flipped
together.

We shall construct the path iteratively, following the direction of the
arrows. Let the bond $b$ be the latest addition to the path. The arrow on
$b$ points to a new vertex $v$. There are two outgoing arrows at $v$,
and what we need is a unique prescription for continuing the path
through $v$. This is provided by a {\em break-up} of the vertex $v$.
In addition to the break-up, we have to allow for {\em freezing} of $v$.
By choosing suitable probabilities for break-up and freezing we shall
satisfy detailed balance.

The {\em break-up} operation is defined by splitting $v$ into two
pieces, as shown in fig.~2.
The two pieces are either two
corners or two straight lines. On each piece, one of the arrows points
towards $v$, while the other one points away from $v$.
Thus we will not allow, e.g., the ul--lr
break-up for a vertex in the configuration 3.
If we break up $v$, the possible new configurations are obtained by
flipping (i.e., reversing both arrows of) the two pieces independently.
On the other hand, if we freeze $v$, the only possible configuration
change is to flip all four arrows.

%
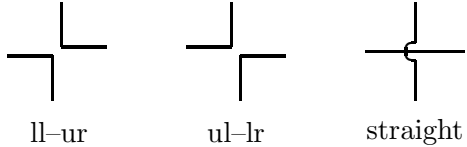
\begin{figure}[tb] \label{figbreakup}
\begin{center}
\setlength{\unitlength}{.0004\textwidth} 
\begin{picture}(1020,210)(0,50) 
\thicklines
\put(100,190){\line(-1, 0){100}}
\put(100,190){\line( 0,-1){100}}
\put(120,210){\line( 1, 0){100}}
\put(120,210){\line( 0, 1){100}}
\put(115,20){\makebox(0,0){ll--ur}}
\put(500,210){\line(-1, 0){100}}
\put(500,210){\line( 0, 1){100}}
\put(520,190){\line( 1, 0){100}}
\put(520,190){\line( 0,-1){100}}
\put(510,20){\makebox(0,0){ul--lr}}
\put(910,200){\line(-1, 0){110}}  
\put(910,200){\line( 1, 0){110}}  
\put(910,180){\line( 0,-1){ 90}}  
\put(910,220){\line( 0, 1){ 90}}  
\put(910,200){\oval(40,40)[l]}    
\put(910,20){\makebox(0,0){straight}}
\end{picture}
\vspace*{1ex}
\caption[fig2]{ \parbox[t]{.80\hsize}{
                The three break-ups of a vertex:
                ll--ur (lower-left--upper-right),
                ul--lr (upper-left--lower-right),
                and straight.
              } }
\end{center}
\vskip-1.5ex   
\end{figure}
%
%

The break-up and freeze probabilities are conveniently described within
the Kandel Domany framework.
It is sufficient to give them for one
vertex, which is in the current configuration $u$.
We define 6 new interactions (weight functions) $\rho_i$, $i=1,...,6$,
corresponding to specific break-up and freeze operations
(the labelling of the new interactions is completely
arbitrary, and the fact that we have six of them is just a coincidence).
For each vertex in configuration $u$, we replace
with probability $p_i(u)$ the original interaction $\rho$ by
the new interaction $\rho_i$.
Equation \ref{KDconstraints}
, i.e.\
detailed balance and the proper normalization of probabilities,
require that for every $u$
\begin{equation}\label{KDforsixvertex}
p_i(u) = q_i \frac{\rho_i(u)}{\rho(u)} \; , \quad \sum_i p_i(u) = 1 \; ,
\end{equation}
where $q_i =\exp{(-c_i)} \geq 0$ are parameters.

\begin{table}[tb]    \label{tabnewinteractions}
\centering
\begin{tabular}{|c|c|c|c|}
\hline\str
 $i$ & action & $\rho_i(\U)$ & $p_i(u)$
\\[.5ex]\hline \stru{4ex}
1 & freeze 1,2 & \parbox[c]{20mm}{1, $\rho(\U)\!=\!a$ \\ 0, else}
               & \parbox[c]{21mm}{$q_1/a$, $\rho(u)\!=\!a$ \\ 0, else}
  \\[2.0ex]\hline \stru{4ex}
2 & freeze 3,4 & \parbox[c]{20mm}{1, $\rho(\U)\!=\!b$ \\ 0, else}
               & \parbox[c]{20mm}{$q_2/b$, $\rho(u)\!=\!b$ \\ 0, else}
  \\[2.0ex]\hline \stru{4ex}
3 & freeze 5,6 & \parbox[c]{20mm}{1, $\rho(\U)\!=\!c$ \\ 0, else}
               & \parbox[c]{20mm}{$q_3/c$, $\rho(u)\!=\!c$ \\ 0, else}
  \\[2.0ex]\hline \stru{4ex}
4 & ll--ur     & \parbox[c]{20mm}{0, $\rho(\U)\!=\!a$ \\ 1, else}
               & \parbox[c]{20mm}{0, $\rho(u)\!=\!a$ \\ $q_4/\rho(u)$, else}
  \\[2.0ex]\hline \stru{4ex}
5 & ul--lr     & \parbox[c]{20mm}{0, $\rho(\U)\!=\!b$ \\ 1, else}
               & \parbox[c]{20mm}{0, $\rho(u)\!=\!b$ \\ $q_5/\rho(u)$, else}
  \\[2.0ex]\hline \stru{4ex}
6 & straight   & \parbox[c]{20mm}{0, $\rho(\U)\!=\!c$ \\ 1, else}
               & \parbox[c]{20mm}{0, $\rho(u)\!=\!c$ \\ $q_6/\rho(u)$, else}
\\[2.0ex] \hline
\end{tabular}
\caption[dummy]{ \parbox[t]{0.85\textwidth}{
                  The new interactions $\rho_i(\U)$
                  and the probabilities $p_i(u)$
                  to choose them at a vertex in current configuration~$u$.
                  See eq.\ (\ref{KDforsixvertex}).
                }}
\end{table}
As discussed in \cite{KandelDomany}
(see also table~1),
{\em freezing} is described by introducing one new interaction for each
different value of $\rho(u)$.
For example, to freeze the value $a$, we choose the
interaction $\rho_1$ to be
$\rho_1(\U)=1$ if $\rho(\U)=a$, and $\rho_1(\U)=0$ otherwise.
In other words,
when $\rho_1$ is chosen,
transitions between $\U=1$ and $\U=2$ cost nothing,
whereas the vertex configurations 3, 4, 5, and 6 are
then not allowed.
Notice that we denote by $u$ the current configuration, and by $\U$
the argument of the function $\rho_i$.

Each {\em break-up} is also described by one new interaction. As an example
take the ul--lr break-up. It is given by the new interaction number five,
with $\rho_5(\U) = 1$ if $\rho(\U)=a$ or $c$,
and $\rho_5(\U) = 0$ if $\rho(\U)=b$.
In other words, with the new interaction $\rho_5$, transitions
between 1, 2, 5 and 6 cost nothing, while the vertex configurations 3
and 4 are not allowed.
This corresponds precisely to allowing independent corner flips
in a ul--lr break-up (see figs.~1,2).

Table~1 gives the full list of new weights $\rho_i(\U)$ and probabilities
$p_i(u)$
to choose them in current configuration $u$.
{}From (\ref{KDforsixvertex}) we also obtain:
\begin{equation}\label{KDnorm}
\begin{array}{l}
q_1+q_5+q_6=a \; , \\
q_2+q_4+q_6=b \; , \\
q_3+q_4+q_5=c \; .
\end{array}
\end{equation}

Assume now that we have broken or frozen all vertices. Starting from a
bond $b_0$, we proceed to construct a closed path by moving in the arrow
direction. As we move from vertex to vertex, we always have a unique way
to continue the path. At broken vertices the path enters the vertex
through one bond and leaves it through another.
If the last bond $b$ added to the cluster points to a frozen vertex $v$,
the path bifurcates in the directions of the two outgoing arrows of $v$.
One of these directions can be considered as
belonging to the loop we came from, the other one as belonging
to a new loop. Since we also have to flip the second incoming
arrow of $v$, we are assured that this new loop also closes.
The two loops have to be flipped together.
In general, the zero-divergence condition
guarantees that all loops will eventually close.


We have now finished describing the procedure for constructing
clusters. In order to specify the algorithm completely, we must choose
values for the constants $q_i$, and decide how the clusters are flipped.
The former problem is of utmost importance, and it is the object of the
next chapter. For the cluster flips, we may use
both the Swendsen-Wang procedure and the Wolff single cluster flip
\cite{ClusterReviews}. We choose the latter, i.e.\ we
``grow'' a {\em single path} from a random starting bond $b_0$, and flip
it. The break-or-freeze decision is only needed for the vertices along
the path, so the computer time for one path is proportional to the
length of that path.

There are some distinct differences between our loop-clusters and more
conventional spin-clusters. For spin-clusters, the elementary objects
that can be flipped are spins; freezing binds them together into
clusters. Our closed loops on the other hand may be viewed as a part of
the {\em boundary} of spin-clusters
(notice that the boundary of spin clusters may
contain loops inside loops). It is reasonable to expect that in
typical cases, building a loop-cluster will cost less work than for a
spin-cluster. This is an intrinsic advantage of the loop algorithm.

\section{Optimization of free parameters}

We have seen that freezing forces loops to be flipped together.
%
%
Previous experience with cluster algorithms 
suggests that it is advantageous to be able to flip loops independently,
as far as possible.
We therefore introduce the principle of {\em minimal freezing}
as a guide for choosing the constants $q_i$:
we shall minimize the freezing probabilities,
given the constraints (\ref{KDnorm}) and $q_i \geq 0$.
In the next chapter we will show that for the case of the F~model, optimization
by minimal freezing does indeed minimize critical slowing down.
Here we discuss optimization for the 4 phases of the 6-vertex model,
usually denoted by capital roman numerals \cite{ReviewLieb}.

Let us first look at phase~IV, where $c > a+b$.
To minimize the freezing of weight $c$, we have to minimize $q_3$.
{}From (\ref{KDnorm}),  $q_3 = c-a-b + q_1+q_2+2q_6$.
With $q_i \geq 0$ this implies $q_{3,\mbox{\small min}} = c-a-b$.
The minimal value of $q_3$ can  only be chosen if {\em at the same time}
we set $q_1=q_2=0$, i.e.\ minimize (in this case do not allow for)
the freezing of the smaller weights $a$ and $b$.
The optimized parameters for phase~IV are then:
\begin{equation} \label{qFOUR}
\begin{array}{lll}
q_1=0, & q_2=0, & q_3=c-a-b, \\
q_4=b, & q_5=a, & q_6=0 \;.
\end{array}
\end{equation}

In phase I the situation is technically similar.
Here $a > b+c$, and we minimize freezing with $q_1=a-b-c$ and $q_2=q_3=0$.
The same holds for phase II, $b>a+c$, where we obtain minimal freezing
for $q_2=b-a-c$ and $q_1=q_3=0$.

Phase III (the massless phase) is characterized by
$a,b,c$ $< \half (a+b+c)$.
Here we can set all freezing probabilities to zero.
Thus,
\begin{equation} \label{qTHREE}
\begin{array}{lll}
q_1=0, & 2q_4=b+c-a \; , \\
q_2=0, & 2q_5=c+a-b \; , \\
q_3=0, & 2q_6=a+b-c \;.
\end{array}
\end{equation}

 \section{Case of the F~model}
%
The F~model is obtained from (\ref{qFOUR}) and (\ref{qTHREE})
as the special case $a=b=\exp{(-K)} \leq 1$, $c=1$.
It has a Kosterlitz-Thouless transition at $K_c=\ln 2$,
with a massless phase for $K\leq K_c$.

How do we choose the parameters $q_i$ here ?
Symmetry $a=b$ implies $q_1=q_2$ and $q_4=q_5$.
We can eliminate freezing of vertices $1,2,3,4,$ by setting
$q_1=q_2=0$. In (\ref{KDnorm}) this leaves one parameter, $q_3$:
\begin{equation}\label{Fmodelparameters}
\begin{array}{lll}
2q_4 &=& 1-q_3 \, , \\
2q_6 &=& \exp{(-K)} +q_3-1 \, .
\end{array}
\end{equation}
In the massless phase, we can minimize freezing by also setting $q_3=0$.
In the massive phase, $q_6\geq 0$ limits $q_3$.
Thus
\begin{equation}\label{qthree}
q_{3,\mbox{\small min}}= \left\{ \begin{array}{ll}
                                   1-2\exp{(-K)}, & K \geq K_c, \\
                                   0,             & K \leq K_c.
                        \end{array}
                \right.
\end{equation}
Notice that since $a=b$ in the F~model, the straight break-up,
the freezing of $a$, and that of $b$ are operationally the same thing.

If we choose $q_3=q_{3,\mbox{\small min}}$,
then for $K \myleq K_c$  vertices of type 5 and 6 are never frozen,
which has as a consequence that every path consists of a single loop.
This loop may intersect itself, like in the drawing of the figure ``8''.
For $K \mygt K_c$ on the other hand,
vertices of type 1, 2, 3 and 4 are never frozen, so we do not
continue a path along a straight line through any vertex.
As $K\rightarrow\infty$ (temperature goes to zero),
most vertices are of type 5 or 6, and they are almost always frozen.
Thus the algorithm basically flips between the two degenerate ground
states.

For the F~model we also have a spin-cluster
algorithm -- the VMR algorithm \cite{BCSOScluster}.
At $K=K_c$ and for $q_3=q_{3,\mbox{\small min}}$,
we  have a situation where the loop-clusters form
parts of the boundary of VMR spin-clusters. Since flipping a
loop-cluster is not the same as flipping a VMR cluster, we expect the
two algorithms to have different performances. We found (see
\cite{BCSOScluster} and the next section) that in
units of clusters, the VMR algorithm is more efficient, but
in work units, which are basically units of CPU time, the loop algorithm
wins. At $K_c/2$, where the loop-clusters are not at all related
to the boundary of VMR clusters,
we found the loop algorithm to be more efficient both in units of
clusters and in work units, with a larger advantage in the latter.

 \section{F-model: Performance of the loop algorithm}
%
We tested the loop algorithm on $L \times L$ square lattices
with periodic boundary conditions at two values of $K$:
at the transition point $K_c$, and
at $\kch$, which is deep inside the massless phase.
%
%
We carefully analyzed autocorrelation functions
and determined the {\em exponential} autocorrelation time $\tau$.
At infinite correlation length, critical slowing down is quantified by
the relation (\ref{CSD}), $\tau\propto L^z$.

Local algorithms are slow, with $z \approx 2$.
For comparison, we performed runs with a local algorithm
that flips arrows around elementary plaquettes with
Metropolis probability, and indeed found $z = 2.2(2)$ at $K=K_c$.

In order to make sure that we do observe the slowest mode of the Markov
matrix, we measured a range of quantities and checked that they exhibit
the same $\tau$.
As in \cite{BCSOScluster}, it turned out that one can use quantities
defined on a sublattice in order to couple strongly to the slowest mode.
Specifically, we wrote the energy as a sum over two sublattice quantities.
We shall present more details of this phenomenon elsewhere.
Let us however note here that for the total energy,
the true value of $\tau$ was not visible within our precision
except for a weak hint of a long tail in the autocorrelations on the
largest lattices we considered.
%
Note that as a consequence of this situation,
the so-called ``integrated  autocorrelation time''
\cite{ClusterReviews} is much smaller than $\tau$, and it would be
completely misleading to evaluate the algorithm based only on its values.
%

We shall quote autocorrelation times $\tau$ in units of
``sweeps'' \cite{ClusterReviews}, defined such that on the average each
bond is updated once during a sweep.
Thus, if $\taucl$ is the autocorrelation time in units of clusters%
, then
$\tau = \taucl \times
       \langle\mbox{cluster size}\rangle / {(2L^2)}$.
Each of our runs consisted of between 50000 and 200000 sweeps.
Let us also define $\zcl$ by
$\taucl \propto L^{\mbox{\protect\small $\zcl$}}$,
and a cluster size exponent $c$ by
$\langle\mbox{cluster size}\rangle \propto L^{\mbox{\protect\small $c$}}$.
We then have:
\begin{equation}\label{taucl}
   z = \zcl - (2-c) \, .
\end{equation}

\begin{table}[tb]
 \centering
\vskip2ex
\begin{tabular}{|r|r|r|}
\hline\str
 $L$ & \mc{1}{c|}{$K=K_c$} & \mc{1}{c|}{$K=\kch$}
\\[.5ex]\hline \str
  8 & 1.8(1)   & 4.9(4) \\
 16 & 3.0(2)   & 5.6(2) \\
 32 & 4.9(4)   & 6.2(3) \\
 64 & 7.2(7)   & 7.4(3) \\
128 &15.5(1.5) & 8.3(2) \\
256 &20.5(2.0) &
\tabhline
 $z$  & \zatKc  &\zatKch
\\[.3ex] \hline
\end{tabular}
 \caption[dummy]{\label{tabtau} \parbox[t]{.85\textwidth}{
                              Exponential autocorrelation time $\tau$
                              at $q_3 \!=\! \qmin$,
                              and the resulting \mbox{dynamical}
                              critical exponent $z$.
                          }}
\end{table}

Table 2 
shows the autocorrelation time $\tau$ for the optimal choice
$q_3\myeq q_{3,\mbox{\small min}}$.
At $K=\kch$, deep inside the massless phase, critical slowing down is almost
completely absent. A fit according to eq.\ \ref{CSD} gives $z=\zatKch$.
The data are also consistent with $z=0$ and only logarithmic growth.
For the cluster size exponent $c$
we obtained $c=1.446(2)$, which points to
the clusters being quite fractal (notice that $\zcl=0.74(2)$).
At the phase transition $K = K_c$ we obtained
$z=\zatKc$, which is still small.
The clusters seem to be less fractal: $c=1.060(2)$,
so that $\zcl=1.65(5)$.

We noted above that at
$K=K_c$ and for the optimal choice of $q_3$, the loop-clusters are related
to the VMR spin-clusters. In \cite{BCSOScluster} we obtained for
the VMR algorithm at $K=K_c$ the result $\zcl=1.22(2)$, but we
had $c=1.985(4)$, which left us with $z=1.20(2)$. In this case, although
in units of clusters the VMR algorithm is more efficient,
the smaller dimensionality of the loop-clusters more than make
up for this, and in CPU time the loop algorithm is more efficient.

As mentioned, no critical slowing down is visible for the integrated
autocorrelation time $\tauint(E)$ of the total energy.
At $K=K_c$, $\tauint(E)$ is only 0.80(2) on the largest lattice,
and we find the dynamical exponent $\zint(E) \approx 0.20(2)$.
At $K=\kch$, $\tauint(E)$ is $1.1(1)$ on all lattice sizes,
so $\zint(E)$ is zero.

What happens for non-minimal values of $q_3\,$?
Table 3 
shows our results on the dependence of $z$ on $q_3$.
$z$ rapidly increases as $q_3$ moves away from $q_{3,\mbox{\small min}}$.
This effect seems to be stronger at $\kch$ than at $K_c$.
We thus see that the optimal value of
$q_3$ indeed produces the best results, as conjectured
{}from our principle of {\em least possible freezing}.

%
\begin{table}[tb]  \centering
\begin{tabular}{|c|c|r|}
\hline\str
 $K$   & $q_3$ & \mc{1}{c|}{$z$}
        \\[.5ex] \hline  \str
$\kch$ & 0    &$ \zatKch$    \\[.3ex]
$\kch$ & 0.10 &$ 1.90(5)$    \\[.3ex]
$\kch$ & 0.20 &$\geq2.6(4)$
\tabhline
$K_c$  & 0     &$ \zatKc $    \\[.3ex]
$K_c$  & 0.05  &$ 0.77(6)$    \\[.3ex]
$K_c$  & 0.10  &$ 0.99(6)$    \\[.3ex]
$K_c$  & 0.20  &$\geq2.2(1)$
\\[.3ex] \hline
\end{tabular}
 \caption[dummy]{\label{tabz}\parbox[t]{.85\textwidth}{
       Dependence of the dynamical critical exponent
       $z$ on the parameter $q_3$.
       We use ``$\geq$'' where for our lattice sizes $\tau$
       increases faster than a power of $L$.
   }}
\end{table}

In the massive phase close to $K_c$,
we expect that
$z(K_c)$ will determine the behaviour of $\tau$
in a similar way as in ref.\ \cite{BCSOScluster}.
To confirm this, a finite size scaling analysis of
$\tau$ is required.
In order to complete the study of the loop algorithm's performance,
we should also investigate it at the KDV transition.

In summary, the loop algorithm strongly reduces critical slowing down,
{}from $z=2.2(2)$ for the Metropolis algorithm,
down to $z=\zatKc$ at $K_c$ and $z=\zatKch$ at $\kch$ deep inside the
massless phase.
For the integrated autocorrelation time of the total energy,
no critical slowing down is visible in either case.
\pagebreak 
 \section{Quantum Spin Systems}

Particularly promising is the possibility of
{\em accelerating Quantum Monte Carlo simulations}, since
quantum spin systems in one and two dimensions can be mapped into
vertex models in $1+1$ and $2+1$ dimensions via the Trotter formula
and suitable splittings of the Hamiltonian \cite{QMC}.
The simplest example is the spin $\half$ $xxz$ quantum chain, which is
mapped directly into the 6-vertex model.
For higher spins, more complicated
vertex models result (e.g.\ 19-vertex model for spin one).

For $(2+1)$ dimensions, different splittings of the Hamiltonian
lead to quite different vertex models, in particular on quite
different lattice types \cite{QMC}. For example, in
the case of spin $\half$ we can choose between 6-vertex models
on a quite complicated $2+1$ dimensional lattice, and models on a
{\em bcc} lattice, with 8 bonds and a large number of configurations
per vertex.

For the simulation of the {\em 2-dimensional Heisenberg antiferromagnet
and ferromagnet} using the former splitting,
all relevant formulas have been worked out
in the present paper.
Actually, the low temperature properties of the antiferromagnet
have recently been investigated by Wiese and Ying
\cite{WieseHeisenberg} {\em using our algorithm}. Their calculation is,
in our opinion, the first high quality verification of the magnon
picture for the low lying excitations.
In particular, this excludes to a much higher degree of confidence
than before the speculation (some years ago widespread) that the
model had a nonzero mass gap.

Notice that, similarly to other cluster algorithms \cite{ClusterReviews},
it is straightforward to define improved observables. The investigation
\cite{WieseHeisenberg} in fact uses them.

Let us also remark that
the loop algorithm can easily change global properties
like the number of world lines or the winding number (see \cite{QMC}).
Thus it is well suited for simulations in the grand canonical ensemble.
Last, but not least, the loop algorithm also opens up a new avenue
for taming the notorious fermion sign problem  \cite{Signpaper}.

%
%

\section{Conclusions}
%
We have presented a new type of cluster algorithm.
It flips closed paths of bonds in vertex models.
Constraints are automatically satisfied.
We have successfully tested our algorithm for the F~model and found
remarkably small dynamical critical exponents.

There are many promising and
straightforward applications of our approach,
to other vertex models, and to 1+1 and 2+1 dimensional
quantum spin systems.
Investigations of such systems are in progress.
In particular, we believe that our generalizations of the
freeze-delete scheme can be adapted for other models like the
8-vertex model.

Already, the loop algorithm has found important applications
in the study of the 2-dimensional Heisenberg antiferromagnet.

\section*{Acknowledgements}

This work was supported in part by the Ger\-man-Israeli
Foundation for Research and Development (GIF) and
by the Basic Research Foundation of
the Israel Academy of Sciences and Humanities.
We would like to express our gratitude
to the HLRZ at KFA J\"ulich 
and to the DOE for providing the necessary computer time
for the F~model study.

   \pagebreak

\end{document}